\documentclass[twocolumn]{aastex631}

\usepackage{multirow}
\usepackage{hyperref}
\usepackage{makecell}
\usepackage{amsmath}
\usepackage{booktabs}
\usepackage{tabularx}
\usepackage{threeparttable}
\usepackage{CJK}

\defcitealias{Rehemtulla+2024_BTSbot}{R24}

\definecolor{DarkOrange}{RGB}{204, 85, 0}
\definecolor{LincolnGreen}{RGB}{17, 102, 0}
\definecolor{Rust}{HTML}{9B4F0F}
\definecolor{DarkCyan}{HTML}{008B8B}
\definecolor{MediumAquaMarine}{HTML}{66CDAA}
\definecolor{Maroon}{HTML}{800000}
\definecolor{Crimson}{HTML}{DC143C}

\newcommand{\revision}[1]{#1}

\shorttitle{Pre-training vision models}
\shortauthors{Rehemtulla et al.}

\begin{document}

\title{Pre-training vision models for the classification of alerts from wide-field time-domain surveys}

\correspondingauthor{Nabeel Rehemtulla}
\email{nabeelr@u.northwestern.edu}

\author[0000-0002-5683-2389]{Nabeel~Rehemtulla}
\affiliation{Department of Physics and Astronomy, Northwestern University, 2145 Sheridan Road, Evanston, IL 60208, USA}
\affiliation{Center for Interdisciplinary Exploration and Research in Astrophysics (CIERA), 1800 Sherman Ave., Evanston, IL 60201, USA}
\affiliation{NSF-Simons AI Institute for the Sky (SkAI), 172 E. Chestnut St., Chicago, IL 60611, USA}

\author[0000-0001-9515-478X]{Adam~A.~Miller}
\affiliation{Department of Physics and Astronomy, Northwestern University, 2145 Sheridan Road, Evanston, IL 60208, USA}
\affiliation{Center for Interdisciplinary Exploration and Research in Astrophysics (CIERA), 1800 Sherman Ave., Evanston, IL 60201, USA}
\affiliation{NSF-Simons AI Institute for the Sky (SkAI), 172 E. Chestnut St., Chicago, IL 60611, USA}

\author[0000-0002-6408-4181]{Mike~Walmsley}
\affiliation{Dunlap Institute for Astronomy \& Astrophysics, University of Toronto, Toronto, ON M5S 3H4, Canada}

\author[0009-0009-1590-2318]{Ved~G.~Shah}
\affiliation{Department of Physics and Astronomy, Northwestern University, 2145 Sheridan Road, Evanston, IL 60208, USA}
\affiliation{Center for Interdisciplinary Exploration and Research in Astrophysics (CIERA), 1800 Sherman Ave., Evanston, IL 60201, USA}
\affiliation{NSF-Simons AI Institute for the Sky (SkAI), 172 E. Chestnut St., Chicago, IL 60611, USA}

\author[0009-0003-6181-4526]{Theophile~Jegou~du~Laz}
\affiliation{Division of Physics, Mathematics, and Astronomy, California Institute of Technology, Pasadena, CA 91125, USA}
\affiliation{NSF Institute on Accelerated AI Algorithms for Data-Driven Discovery (A3D3)}

\author[0000-0002-8262-2924]{Michael~W.~Coughlin} 
\affiliation{NSF Institute on Accelerated AI Algorithms for Data-Driven Discovery (A3D3)}
\affiliation{School of Physics and Astronomy, University of Minnesota, Minneapolis, MN 55455, USA}

\author[0000-0001-7357-0889]{Argyro Sasli}
\affiliation{NSF Institute on Accelerated AI Algorithms for Data-Driven Discovery (A3D3)}
\affiliation{School of Physics and Astronomy, University of Minnesota, Minneapolis, MN 55455, USA}

\author[0000-0002-7777-216X]{Joshua~Bloom}
\affiliation{Physics Division, Lawrence Berkeley National Laboratory, 1 Cyclotron Road, Berkeley, CA, 94720, US}
\affiliation{Department of Astrophysics, University of California, Berkeley, CA
94720-3411, USA}

\author[0000-0002-4223-103X]{Christoffer Fremling}
\affiliation{Division of Physics, Mathematics, and Astronomy, California Institute of Technology, Pasadena, CA 91125, USA}
\affiliation{Caltech Optical Observatories, California Institute of Technology, Pasadena, CA 91125, USA}

\author[0000-0002-3168-0139]{Matthew~J.~Graham}
\affiliation{Division of Physics, Mathematics, and Astronomy, California Institute of Technology, Pasadena, CA 91125, USA}

\author[0000-0001-5668-3507]{Steven L. Groom}
\affiliation{IPAC, California Institute of Technology, 1200 E. California Blvd, Pasadena, CA 91125, USA}

\author{David~Hale}
\affiliation{Caltech Optical Observatories, California Institute of Technology, Pasadena, CA 91125, USA}

\author[0000-0003-2242-0244]{Ashish~A.~Mahabal}
\affiliation{Division of Physics, Mathematics, and Astronomy, California Institute of Technology, Pasadena, CA 91125, USA}
\affiliation{Center for Data Driven Discovery, California Institute of Technology, Pasadena, CA 91125, USA}

\author[0000-0001-8472-1996]{Daniel A. Perley}
\affiliation{Astrophysics Research Institute, Liverpool John Moores University, 146 Brownlow Hill, Liverpool L3 5RF, UK}

\author[0000-0003-1227-3738]{Josiah~Purdum}
\affiliation{Caltech Optical Observatories, California Institute of Technology, Pasadena, CA 91125, USA}

\author[0000-0001-7648-4142]{Ben Rusholme}
\affiliation{IPAC, California Institute of Technology, 1200 E. California Blvd, Pasadena, CA 91125, USA}

\author[0000-0003-1546-6615]{Jesper Sollerman}
\affiliation{Department of Astronomy, The Oskar Klein Center, Stockholm University, AlbaNova, SE-10691 Stockholm, Sweden}

\author[0000-0002-5619-4938]{Mansi~M.~Kasliwal}
\affiliation{Division of Physics, Mathematics, and Astronomy, California Institute of Technology, Pasadena, CA 91125, USA}

\begin{abstract}

Modern wide-field time-domain surveys facilitate the study of transient, variable and moving phenomena by conducting image differencing and relaying alerts to their communities. Machine learning tools have been used on data from these surveys and their precursors for more than a decade, and convolutional neural networks (CNNs), which make predictions directly from input images, saw particularly broad adoption through the 2010s. Since then, continually rapid advances in computer vision have transformed the standard practices around using such models. It is now commonplace to use standardized architectures pre-trained on large corpora of everyday images (e.g., ImageNet). In contrast, time-domain astronomy studies still typically design custom CNN architectures and train them from scratch. Here, we explore the \revision{effects} of adopting various pre-training regimens and standardized model architectures on the performance of alert classification. We find that the resulting models match or outperform a custom, specialized CNN like what is typically used for filtering alerts. Moreover, our results show that pre-training on galaxy images from Galaxy Zoo tends to yield better performance than pre-training on ImageNet or training from scratch. We observe that the design of standardized architectures are much better optimized than the custom CNN baseline, requiring significantly less time and memory for inference despite having more trainable parameters. On the eve of the Legacy Survey of Space and Time and other image-differencing surveys, these findings advocate for a paradigm shift in the creation of vision models for alerts, demonstrating that greater performance and efficiency, in time and in data, can be achieved by adopting the latest practices from the computer vision field.

\end{abstract}


\section{Introduction} \label{sec:intro}
Large, modern, wide-field time-domain surveys like the Zwicky Transient Facility (ZTF; \citealt{Bellm+2019a, Bellm+2019b, Graham+2019, Masci+2019, Dekany+2020}), the Asteroid Terrestrial-impact Last-Alert System \citep[ATLAS;][]{Tonry+2011, Tonry+2018, Smith+2020}, the Panoramic Survey Telescope and Rapid Response System \citep[Pan-STARRS;][]{Kaiser+2002} and the Vera C. Rubin Observatory's Legacy Survey of Space and Time (LSST; \citealp{Ivezic+2019}) employ image differencing to identify new, moving, and varying sources in real-time. Image differencing involves subtracting a reference image from a new science image and searching for statistically significant source detections in the resulting difference image. These detections create alerts which are promptly relayed to the community via an alert stream \citep[e.g.,][]{Patterson+2019} through alert brokers (e.g., \citealp{Duev+2019, Forster+2021, antares_matheson21, Moller+2021, Williams+2024, JegouduLaz+2025}), and they typically include image cutouts of the new, reference, and difference images centered on the source location. Alert streams are fertile ground for studying an array of astrophysical phenomena, including variable stars (VarStars), galactic and extragalactic transients, active galactic nuclei (AGN), and asteroids. Consequently, significant filtering of an alert stream is required to isolate events relevant to specific scientific cases.
See \cite{Rehemtulla+2025b_NatAstP} for an overview of the current state of these workflows and a projection for how they will evolve in the Rubin era.

\begin{figure*}[ht!]
    \centering
    \includegraphics[width=1\linewidth]{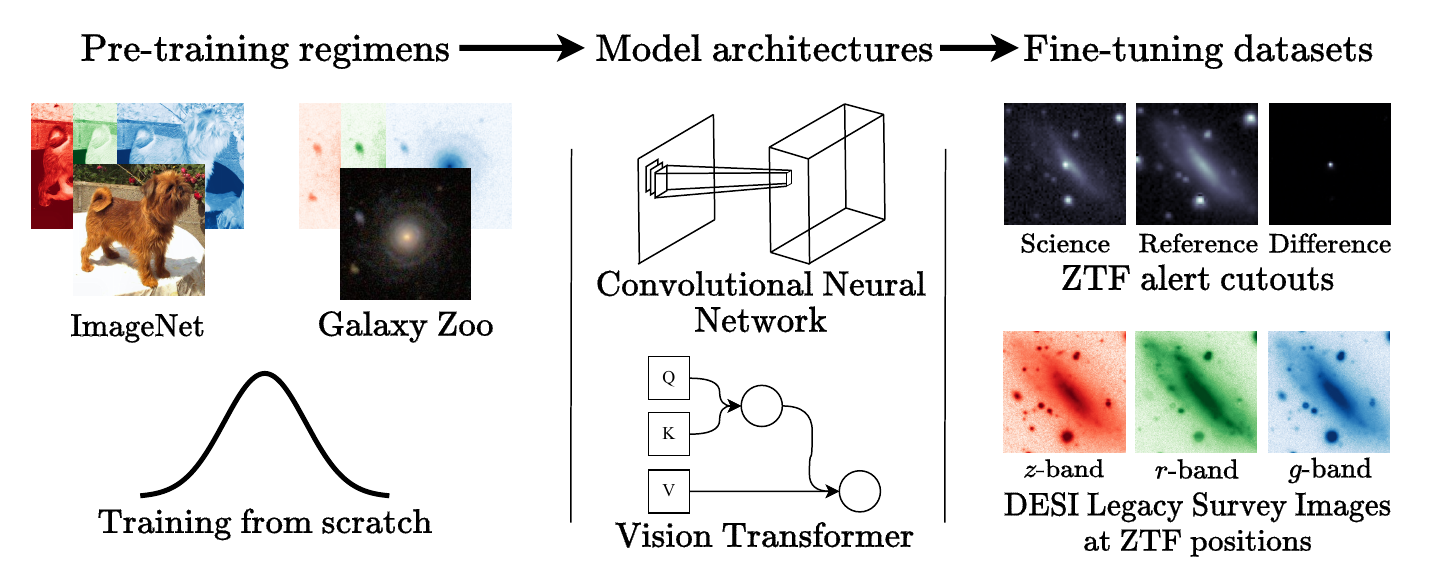}
    \caption{Schematic of our experimental design. We evaluate model performance for all permutations of the depicted pre-training regimens, model architectures, fine-tuning datasets, as well as four settings of fine-tuning dataset size.}
    \label{fig:diagram}
\end{figure*}

There exists a rich legacy of machine learning (ML) classifiers enabling time-domain astronomy by assisting with filtering alert streams. Early work adopted models such as support vector machines and tree-based classifiers \citep[e.g.,][]{Bailey+2007, Bloom+2012, Brink+2013} which produced classifications using descriptive statistical features extracted from the data. Later, the extraordinarily rapid development of computer vision techniques spurred the adoption of convolutional neural networks \citep[CNNs;][]{Fukushima+1982, LeCun+2015} which eliminate the need for manually designing statistical features to extract from the data by processing image cutouts directly and often yielding fruitful models \citep[e.g.,][]{Gieseke+2017, Cabrera-Vives+2017, Sedaghat+2018, Duev+2019, Duev+2021b, Duev+2021, Carrasco-Davis+2021, Sheng+2024, Weston+2024, Rehemtulla+2024_BTSbot}. 

Continued rapid progress in computer vision, supported by large, labeled, diverse and publicly available datasets provided the scale and breadth necessary for training modern vision models. MNIST \citep{Deng_2012} and CIFAR \citep{Krizhevsky+2009} standardized evaluation, while the introduction of large-scale datasets like ImageNet \citep{Russakovsky+2014}, COCO \citep{Lin+2014} and Places \citep{Zhou+2017} provided data diversity through immense scale. The standard practices around creating computer vision models have evolved with time, in part to take advantage of the availability of these datasets. It is now commonplace in computer vision studies to adopt pre-trained models with standardized, or ``off-the-shelf," architectures for practically any task, often irrespective of whether or not the pre-training dataset is closely related to the task at hand. Pre-training is the process of training a model on a large, generic dataset to learn general feature extraction capabilities before conducting a task-specific training routine (``fine-tuning"). Pre-training confers benefits including faster convergence, stronger generalization, and lower data requirements \citep{Girshick+2013, He+2021}. Alongside this, off-the-shelf architectures offer well-tested baselines with proven design choices, allowing researchers to avoid time-consuming low-level model design. 

Some time-domain astronomy studies creating vision models follow these practices and have deployed their products into production workflows \citep{Weston+2024, Junell+2025} but most still train custom CNNs from scratch. Those that do follow modern practices, e.g. \texttt{AppleCiDEr} \citep{Junell+2025}, also tend to include analyses demonstrating the superior performance and/or computational efficiency offered by pre-trained off-the-shelf vision models. 
\cite{Stoppa+2025} take an entirely different approach and explore real-bogus classification of alert cutouts with a multi-modal large language model (LLM). The integration of an LLM allows the for the model to provide textual explanations of its classifications, although the considerable computational cost of these models is not conducive to real-time deployment. Studies in other areas of astronomy have more broadly adopted pre-training and off-the-shelf architectures for their vision models \citep[e.g.,][]{Wu-Boada-2019, Walmsley+2022, VafaeiSadr+2023, Mohale-Lochner-2024}. Researchers creating models to operate on time-series data have also taken a special interest in pre-training, in part to overcome small training sets typically with large class imbalances \citep{Fiscale+2021, Zhang+2024, Gupta+2025a, Gupta+2025b}. These works have demonstrated that pre-training and off-the-shelf architectures can unlock models that yield better performance with less, or even zero \citep{Li+2025}, task-specific training data. 

In this work, we explore a wide breadth of design choices faced when creating vision models for alert classification. Our experiments measure classification performance on a transient candidate vetting task with two pre-training regimens (Section~\ref{sec:choice-pretraining}) compared against training from scratch, two standardized model architectures with orthogonal philosophies (Section~\ref{sec:choice-arch}), and two systematically different fine-tuning datasets (Section~\ref{sec:choice-btsbot}). These experiments are also repeated across an order of magnitude difference in training set size to reveal trends as a function of the quantity of labeled data available for training. Although custom CNNs trained from scratch have served time-domain astronomers well for years, the growth of our data quantity and variety in the Rubin-era will challenge this paradigm because it is expected to far outpace the growth of data labeling. Such an investigation is particularly timely as there are numerous image-differencing facilities which have recently—or will soon—become operational: LSST, the BlackGEM Telescope Array \citep{Groot+2024}, the Nancy Grace Roman Space Telescope \citep{Akeson+2019}, the La Silla Schmidt Southern Survey \citep[LS4;][]{Miller+2025}, the Argus Optical Array \citep{Law+2022}, and others. We conclude with a review of our findings which serve as a set of recommendations for most efficiently creating vision models to be deployed into the alert streams of these current and upcoming image-differencing surveys.

\section{Experimental design}
\label{sec:exp-design}

As shown in Figure~\ref{fig:diagram}, we explore vision model design by evaluating models using various off-the-shelf architectures (Section~\ref{sec:choice-arch}), pre-training regimens (Section~\ref{sec:choice-pretraining}) and fine-tuning datasets on a transient candidate vetting task (Section~\ref{sec:choice-btsbot}). We train models with modern optimization techniques and hyperparameter sweeps (Section~\ref{sec:hpo}) and present results as a function of training set size.

\subsection{Choice of off-the-shelf architectures}
\label{sec:choice-arch}

We select the ConvNeXt \citep{Liu+2022} and MaxViT \citep{Tu+2022} architectures to evaluate against a small, custom, specialized CNN like those often used in the time-domain astronomy literature. This choice reflects the aim to compare two dominant classes of modern but fundamentally different vision models: fully-convolutional models (i.e., those without attention mechanisms) and vision transformers \citep[ViT;][]{Dosovitskiy+2020}, i.e., those predominantly driven by attention mechanisms.

ConvNeXt \citep{Liu+2022} represents a modern reinterpretation of CNNs by revisiting the ResNet architecture \citep{He+2016} and borrowing insights from recent transformer architectures \citep{Vaswani+2017} as well as other highly efficient CNNs \citep{Chollet+2016, Howard+2017}.
In general, CNNs apply small, sliding filters across an image to detect local patterns, textures, or small-scale structures. \revision{Numerous studies have shown that CNNs can very effectively capture real astronomical features like galaxy morphologies \citep[e.g., ][]{Bhambra+2022, Alfonzo+2024}.} The ConvNeXt architecture preserves the strong inductive bias toward translational invariance and feature locality inherent to CNNs while incorporating architectural refinements such as large kernel sizes and improved normalization layers.
\revision{\citep{Liu+2022} present detailed ablation studies and interpretation to why certain choices improve performance.}
Overall, these updates allow ConvNeXt models to achieve performance competitive with ViTs on large-scale image classification benchmarks, while retaining the stability and computational efficiency characteristic of CNNs \citep{Liu+2022}.

MaxViT \citep{Tu+2022} represents a refinement of the original ViT design. ViTs use attention mechanisms to learn relationships across an entire image and to prioritize regions of particular interest.
\revision{While the CNN's convolution operations only consider a region the size of a single kernel at a time, the ViT's self-attention mechanism allows for each input element to ``attend'' to each other element. In practice, these input elements are tokens which originate from patches of pixels and are then fed to the attention heads. This balances the substantial computational cost of the attention layers while maintaining its flexibility and generality. Stemming from the immense representation capabilities of ViTs, they have become established as a highly performant tool across a breadth of domains \citep{Khan+2022}.}
The MaxViT architecture introduces a ``multi-axis" attention mechanism which alternates between local and global self-attention, allowing the model to capture relationships across large and small scales without a steep increase in computational cost.

Together, ConvNeXt and MaxViT represent complementary points along the inductive bias spectrum, enabling an assessment of whether convolutions or attention mechanisms yield superior performance and efficiency.

Our model architecture choices are affirmed by the large architecture benchmarking experiment presented by \cite{Walmsley+2024}. In their study, performance on a galaxy morphology task is measured for various modern architectures, model sizes, and training dataset sizes. They find that ConvNeXt and MaxViT perform the best overall, and that the larger, higher parameter count versions of the ConvNeXt and MaxViT models tend to only outperform their smaller analogs when accompanied by significantly larger quantities of training data. Given the limited size $(N \approx25,000$ total sources) of our training set (Section~\ref{sec:choice-btsbot}), we select the smallest available versions of the ConvNeXt and MaxViT architectures: ConveNeXt-pico (8.5M trainable parameters) and MaxViT-tiny (28.5M trainable parameters).\footnote{Number of trainable parameters are counted only from the model's vision backbone and exclude any trainable parameters in the model's MLP head.}

When training any model, its head is replaced by a randomly initialized multi-layer perceptron (MLP) with three layers whose hyperparameters are optimized with the procedure described in Section~\ref{sec:hpo}. During training, all weights and biases in the vision backbone and the MLP head are unfrozen and allowed to be updated.

\subsection{Choice of pre-training regimens}
\label{sec:choice-pretraining}

We select two pre-training regimens to compare against training from scratch: pre-training on ImageNet \citep{Russakovsky+2014} and pre-training on Galaxy Zoo \citep{Lintott+2008, Lintott+2011}.
Comparing results across these configurations allows us to asses the effect of pre-training relative to training from scratch and the importance of pre-training on real astronomical observations. \revision{In other terms, we hope to assess whether pre-training on real astronomical observations (Galaxy Zoo) yields better results than naïve pre-training on entirely out-of-domain non-astronomy data (ImageNet) and if pre-training on out-of-domain data provides anything useful relative to training from scratch. These goals are deliberately oriented towards providing time domain computer vision practitioners with useful guidance when creating models for use with alert cutouts.}

ImageNet \citep{Russakovsky+2014} is a large, labeled dataset containing images of everyday objects, natural scenes, animals, etc. while excluding astronomical observations.
In this study we adopt the ImageNet-1k subset which contains $\sim$1.28 million images across 1,000 classes.
We select it here because it is a literature-standard choice for benchmarking vision models.
We obtain the ConvNeXt and MaxViT models pre-trained on ImageNet-1k from the PyTorch Image Models library \citep[\texttt{timm};][]{timm}.

Galaxy Zoo \citep{Lintott+2008, Lintott+2011} is a citizen science project where volunteers classify galaxies in real astronomical images based on their morphologies. The Galaxy Zoo project began with volunteers annotating images from the Sloan Digital Sky Survey \citep{York+2000} and has since expanded to images from the Hubble Space Telescope, the Dark Energy Spectroscopic Instrument's (DESI's) Legacy Survey \citep{DESI+2016, Dey+2019}, the UKIRT Infrared Deep Sky Survey \citep[UKIDSS;][]{Lawrence+2007}, and more. Over nearly two decades, $>$100 thousand Galaxy Zoo volunteers have contributed $>$100 million total annotations. In this study we use \texttt{Zoobot} models pre-trained by \cite{Walmsley+2024} on $\sim$842,000 annotated Galaxy Zoo images from five observatories. The \texttt{Zoobot} models learn galaxy morphology by being trained to predict the volunteers' answers to questions posed by Galaxy Zoo (see \citealt{Walmsley+2024} for details). We obtain the models pre-trained by \cite{Walmsley+2024} from the \texttt{Hugging Face} platform.\footnote{\url{https://huggingface.co/mwalmsley}}

We compare these two pre-training regimens against training from scratch, which involves drawing values for the model's trainable parameters from distributions carefully designed to stably and quickly being the training process.
In all cases, we adopt the default initialization schemes used in PyTorch \citep{Paszke+2019}, which make use of findings from studies exploring initialization strategies \citep[e.g.,][]{GlorotBengio2010, He+2015}.
Models without pre-training are obtained from \texttt{timm}.

\subsection{Choice of training set and task}
\label{sec:choice-btsbot}

We adopt the \texttt{BTSbot} \citep{Rehemtulla+2024_BTSbot} training set and transient candidate vetting task because they provide high quality labels on a very large training set with representation across a wide range of source types. \texttt{BTSbot} originates from the ZTF Bright Transient Survey \citep[BTS;][]{Fremling+2020, Perley+2020, Rehemtulla+2024_BTSbot}, an effort aiming to spectroscopically classify all bright ($m_{\textrm{peak}}\leq18.5$ mag in $g_\textrm{ZTF}$ or $r_\textrm{ZTF}$) extragalactic transients in the ZTF public alert stream. BTS members conduct daily visual inspection of bright transient candidates and send promising events to the SED Machine \citep{Blagorodnova+2018, Rigault+2019, Kim+2022} and other spectrographs for classification. Over its $\sim$7\,yr lifetime, BTS has produced an immense dataset with high quality binary labels. Still, there is inevitably non-zero label noise. In Section~\ref{sec:further_comparison}, we do identify two mislabeled sources, but we expect that the level of label noise is not significant enough to have a substantial detrimental \revision{effect} on the performance of models we train here. 

\cite{Rehemtulla+2024_BTSbot} compiled sources selected and rejected from the BTS sample to train \texttt{BTSbot} to automate bright transient candidate vetting. This task involves considering a wide breadth of sources including VarStars, AGN, asteroids, galactic and extragalactic transients, and even alerts produced by artifacts of the image differencing process (``bogus" alerts). The comprehensive nature of this training set and task make it an appropriate choice for representatively probing the \revision{effect} of pre-training on alert classification models. Moreover, the \texttt{BTSbot} training set is significantly larger than that of other comparable, published models, allowing us to experiment with pre-training and model architectures at greater scales. We rerun the training set compilation and pre-processing described by \cite{Rehemtulla+2024_BTSbot} which makes use of ZTF's first-party alert broker \citep[Kowalski;][]{Duev+2019} and marshal \citep[SkyPortal/Fritz;][]{van_der_Walt+2019, Coughlin+2023} to update the \texttt{BTSbot} training set originally queried in August 2023. The updated training set contains 769,056 alerts across 25,609 sources: $\sim$30\% larger than the original.


The ZTF alert cutouts in this dataset utilize a significantly different arrangement than those in the pre-training datasets.
Both ImageNet and Galaxy Zoo have three-channel red-green-blue images while the three channels of the ZTF alert cutouts represent the science, reference and difference images (see right side of Figure~\ref{fig:diagram}).
\revision{The red-green-blue images naturally stack into a color composite image where the relative pixel values of a given region across channels conveys additional astrophysical information, e.g. bluer parts of a galaxy tend to be populated with young, massive, hot stars and thus are associated with recent star formation. Alert cutouts produced by image differencing do not share this property; instead, the difference image (third channel) is a linear combination of the science and reference images (first and second channels). Instead of color information, this schema offers information on the quality of the subtraction and the morphology of the source's PSF.}
Thus, the cross-channel correlations between these two image schema are dramatically different. This introduces a distribution shift between the pre-training and the fine-tuning data.
In order to probe the consequence of this difference, we introduce an additional fine-tuning dataset consisting of DESI Legacy Survey data release 10 (DR10) $grz$ color images at the same field-of-view (63"$\times$63") and pixel-scale (1" per pixel) as the ZTF alert cutouts.\footnote{This is not the native pixel-scale of Legacy Survey images, however, we bin the images to this pixel-scale to match that of the ZTF alert cutouts.} Thus, this Legacy Survey dataset contains the same environmental contextual information critical for transient candidate vetting while minimizing the distribution shift between pre-training and fine-tuning. Legacy Survey also provides greater depth and introduces color information, both of which are absent in the ZTF alert cutouts. A significant subset of alerts ($\sim$22\%) are omitted from this fine-tuning dataset because there is no Legacy Survey DR10 coverage at their position. The omitted alerts predominantly occur at low galactic latitudes, so certain phenomena, like cataclysmic variables, are disproportionately removed relative to others. Thus, an absolute comparison of model performance metrics between the two fine-tuning datasets is flawed but comparing the relative performance differences between pre-training regimens can aid with better understanding the distribution shift between the ZTF alert cutouts and the pre-training datasets.

We also assess the data efficiency of each model configuration by repeating experiments for four settings of training set size. We create smaller versions of the train split by randomly selecting 5, 10, and 50\% of alerts to preserve while removing all others. The models trained on the smaller datasets still use the full validation and test splits to ensure performance metrics are representative of the model's true performance.

Both the ConvNeXt and MaxViT models we select take in images much larger than that of the image cutouts from either fine-tuning dataset. To handle this, we up-sample our input images to the model's expected resolution \revision{while maintaining smooth source profiles with bilinear pixel interpolation.}

\subsection{Training and hyperparameter optimization}
\label{sec:hpo}

\begin{deluxetable}{cc}
\tablecaption{Hyperparameter optimization \label{tab:hpo}}
\tablehead{
    \colhead{Hyperparameter name} & \colhead{Sweep Search Range}
}
\startdata
    \multicolumn{2}{c}{Optimizer} \\
    \hline
    Learning rate ($\alpha$) & $\{1,5\}\times10^{-6}, \{1,5\}\times10^{-7}$ \\
    Warm-up Epochs & $\{2,4,6\}$ epochs \\
    Batch size & $\{16, 64, 256\}$ \\
    \hline
    \multicolumn{2}{c}{Standardized MLP head} \\
    \hline
    Dense layer 1 & $\{128, 256, 512\}$ units \\
    Dropout layer & $0.3-0.9$ \\
    Dense layer 2 & $\{8, 16, 32\}$ units \\
    \hline
\enddata
\end{deluxetable}

Models are trained with an updated version of the \texttt{BTSbot} training code\footnote{\url{https://github.com/nabeelre/BTSbot}} rewritten to use the PyTorch API \citep{Paszke+2019}. Most aspects of training are kept consistent with that of \texttt{BTSbot}, e.g., the binary cross entropy loss function, data augmentation techniques and class weighting \citep{Rehemtulla+2024_BTSbot}. Some select settings are updated to more modern practices; we adopt the \texttt{AdamW} optimizer \citep{Loshchilov+2017} and a learning rate scheduler which contains a linear warm-up followed by a cosine decay \citep{Loshchilov+2016}. The best performing optimizer and model hyperparameters can vary for each selection of pre-training regimen, architecture, and fine-tuning dataset, so we execute a separate Bayesian hyperparameter sweep for each configuration to ensure they are fairly assessed. Table~\ref{tab:hpo} shows the hyperparameters ranges over which we search. Sweeps are conducted using the Weights and Biases platform \citep{wandb}, and we find that $<60$ trials per sweep is typically sufficient to converge to a near-optimal set of hyperparameters.\footnote{We do identify a small number of cases where this number of trials is insufficient and the measured performance slightly diverts from trends observed in related sweeps (see Secs.~\ref{sec:results-sm}, \ref{sec:results-mm}).} Once a sweep is completed, we use the hyperparameters of the best performing trial to run five final trials with the optimized hyperparameters but with varied random seeds. The performance metrics we report are the median and standard deviation for these final trials computed on the test split.

We primarily focus on models trained only with images to distill the \revision{effect} of the updated vision model design, but we also explore multi-modal models to serve as a direct replacement to the production \texttt{BTSbot} model. Models are made multi-modal by introducing a metadata branch whose embedding is concatenated with that of the image branch before being passed to an MLP head. The metadata features are taken exactly from that of the production \texttt{BTSbot} model.

Lastly, we also re-train the uni-modal and multi-modal versions of the custom CNN used by \texttt{BTSbot} to serve as a comparison baseline. Many other models performing alert classification use similar architectures \citep[e.g.,][]{Andreoni+2017, Gieseke+2017, Duev+2019, Duev+2021b, Carrasco-Davis+2021, Duev+2021, Makhlouf+2022, Acero-Cuellar+2023, Sheng+2024, Weston+2024, Liu+2025, Pranshu+2025}, so this baseline represents the most common approach in the literature. Similarly to the off-the-shelf architectures, we execute hyperparameter sweeps using the search ranges described in Table~\ref{tab:hpo} and in \cite{Rehemtulla+2024_BTSbot}.

\section{Results from single-modality models} \label{sec:results-sm}

\begin{figure*}[ht!]
    \centering
    \includegraphics[width=0.495\linewidth]{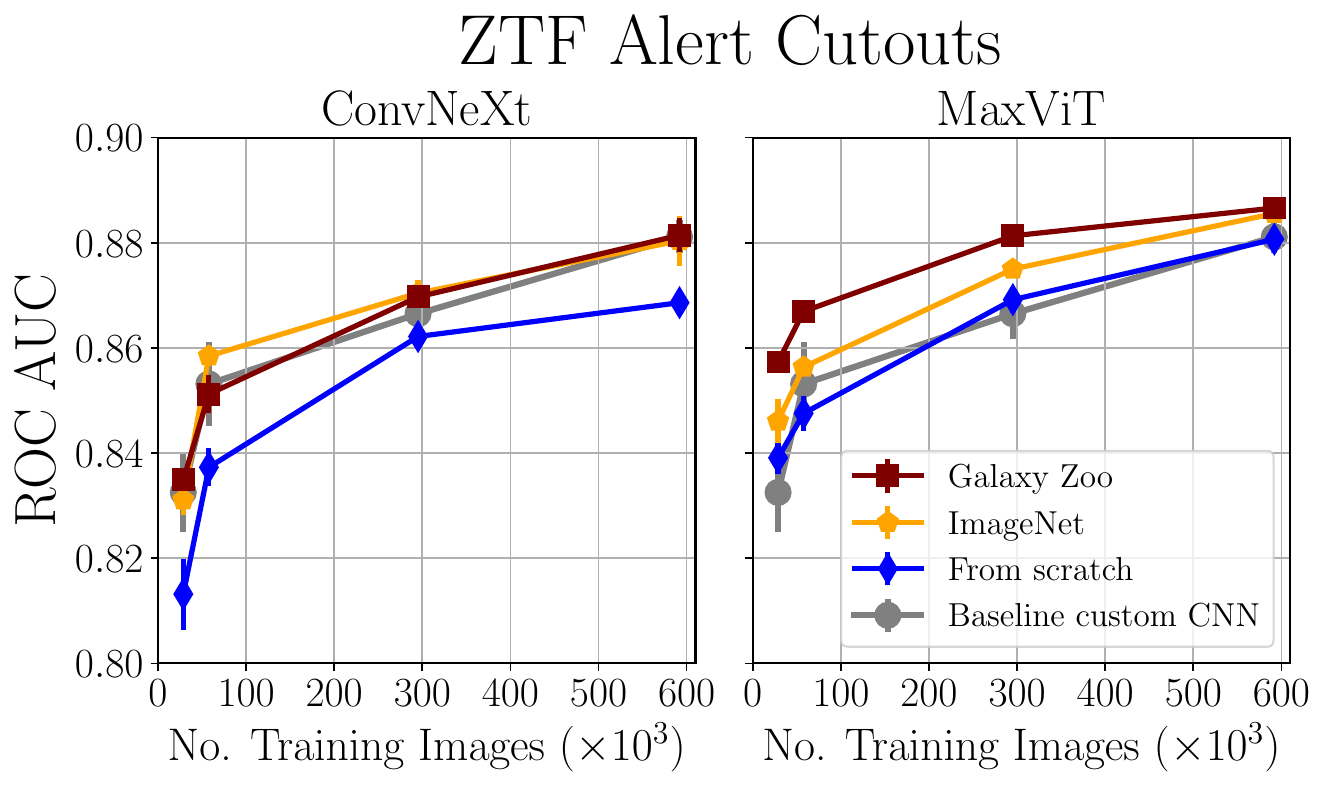}
    \includegraphics[width=0.495\linewidth]{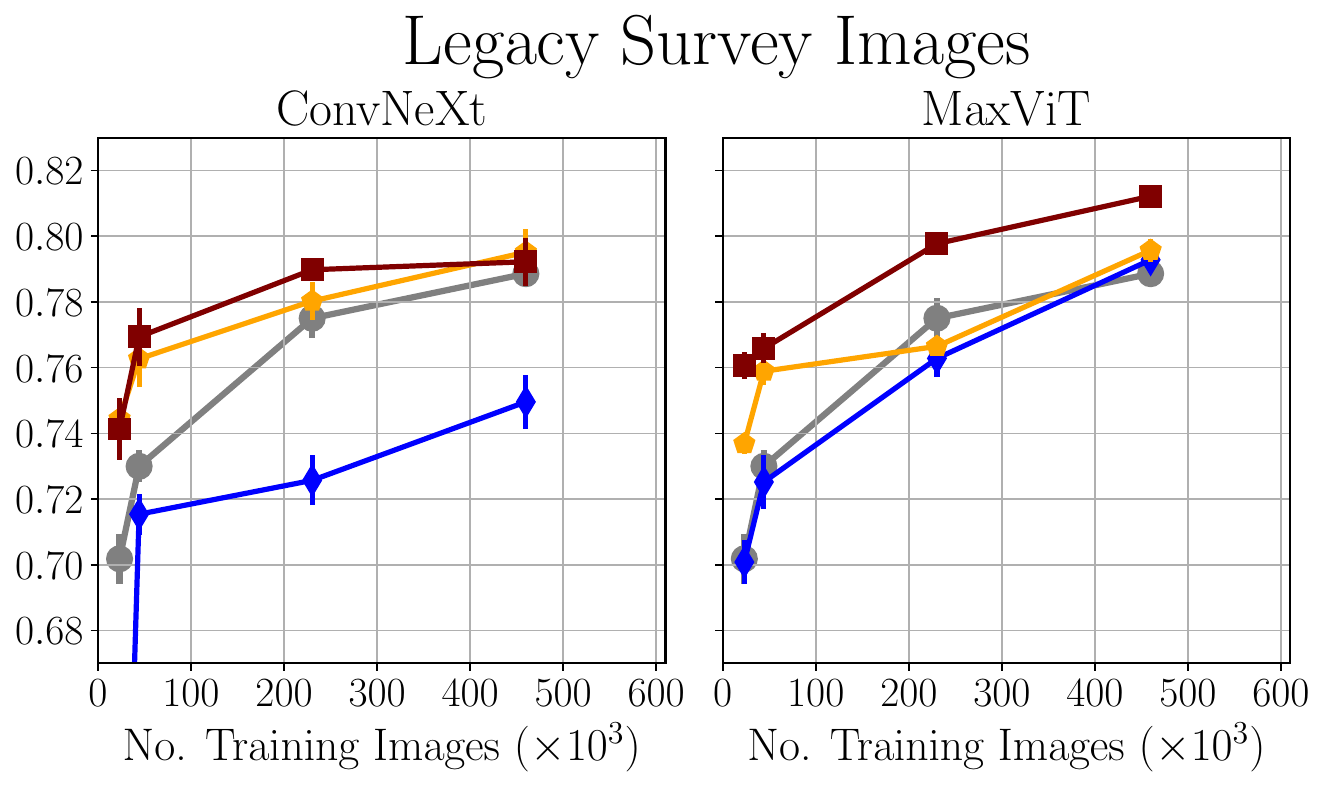}
    \caption{Area under the receiver operating characteristic curve (ROC AUC) for experiments varying the model's pre-training regimen, architecture, and fine-tuning dataset. Comparisons are made against a custom architecture CNN (gray) alike those often adopted in the astronomy literature. \textit{Left:} Models fine-tuned on ZTF alert cutouts. Across both the ConvNeXt and MaxViT architectures, pre-training on Galaxy Zoo (maroon) or ImageNet (orange) yields better performance than training from scratch (blue). The Galaxy Zoo MaxViT outperforms the baseline and the ImageNet MaxViT at all fine-tuning dataset sizes. \textit{Right:} Models fine-tuned on Legacy Survey images. Pre-trained models usually significantly outperform their counterparts without pre-training. Again, the Galaxy Zoo MaxViT model surpasses the performance of all others.}
    \label{fig:images-only}
\end{figure*}

We present results of models benchmarked on bright transient candidate vetting. Model performance is evaluated using the area under the receiver operating characteristic curve (ROC AUC), a general-purpose indicator of classification performance that accounts for the mild class imbalance present in the training set. Higher values of ROC AUC indicate better classification performance, reaching unity for perfect classification. 

Figure~\ref{fig:images-only} reports ROC AUC as a function of the number of images trained on. For the ZTF alerts dataset, we find that the pre-trained ConvNeXt models yield performance consistent within uncertainties with the baseline custom CNN across all training set sizes. The ConvNeXt models trained from scratch, however, slightly underperform their pre-trained analogs and the baseline CNN in all tests: they yield ROC AUC score up to $\sim$0.02 lower than the baseline. The MaxViT models trained from scratch closely trace the performance of the baseline, and the models pre-trained on ImageNet show a slight boost in performance relative to those without pre-training. The Galaxy Zoo MaxViT models yield the best performance across all training set sizes, reaching a maximum of $\sim$0.025 higher ROC AUC than the baseline. Their improvement relative to ImageNet pre-training (a median of $\sim$0.01 in ROC AUC) suggests that some aspect of the galaxy images promoted transfer to the ZTF alert cutouts more readily than the seemingly unrelated content in the ImageNet images. 

We also experiment with fine-tuning the off-the-shelf architectures on Legacy Survey cutouts to test the differential effect of pre-training on their red-green-blue ($zrg$ bands) layout versus the science-reference-difference layout of the ZTF alert cutouts. The right panels of Figure~\ref{fig:images-only} shows that the ConvNeXt models fine-tuned on Legacy Survey images unambiguously illustrate a strong advantage toward the pre-trained models relative to training from scratch. In these experiments, the relative performance gain from pre-training ($\gtrsim0.04$) is much larger than in the experiments fine-tuning on ZTF alert cutouts ($\lesssim0.01$). We posit that this is a product of the reduced data shift between the fine-tuning and pre-training datasets. Similarly to their analogs fine-tuned on ZTF images, the MaxViT models fine-tuned on Legacy Survey images show that pre-training on Galaxy Zoo yields a clear advantage relative to pre-training on ImageNet and training from scratch. As in the ZTF alert cutouts test, the MaxViT model pre-trained on Galaxy Zoo yields the best overall performance. We note that the performance of the Legacy Survey models is systematically worse than that of the models operating on ZTF alert cutouts. We do not draw any conclusions from this result as it does not represent a level comparison (see Section~\ref{sec:choice-btsbot}).

\section{Results from multi-modal models} \label{sec:results-mm}

\begin{figure*}[ht!]
    \centering
    \includegraphics[width=0.55\linewidth]{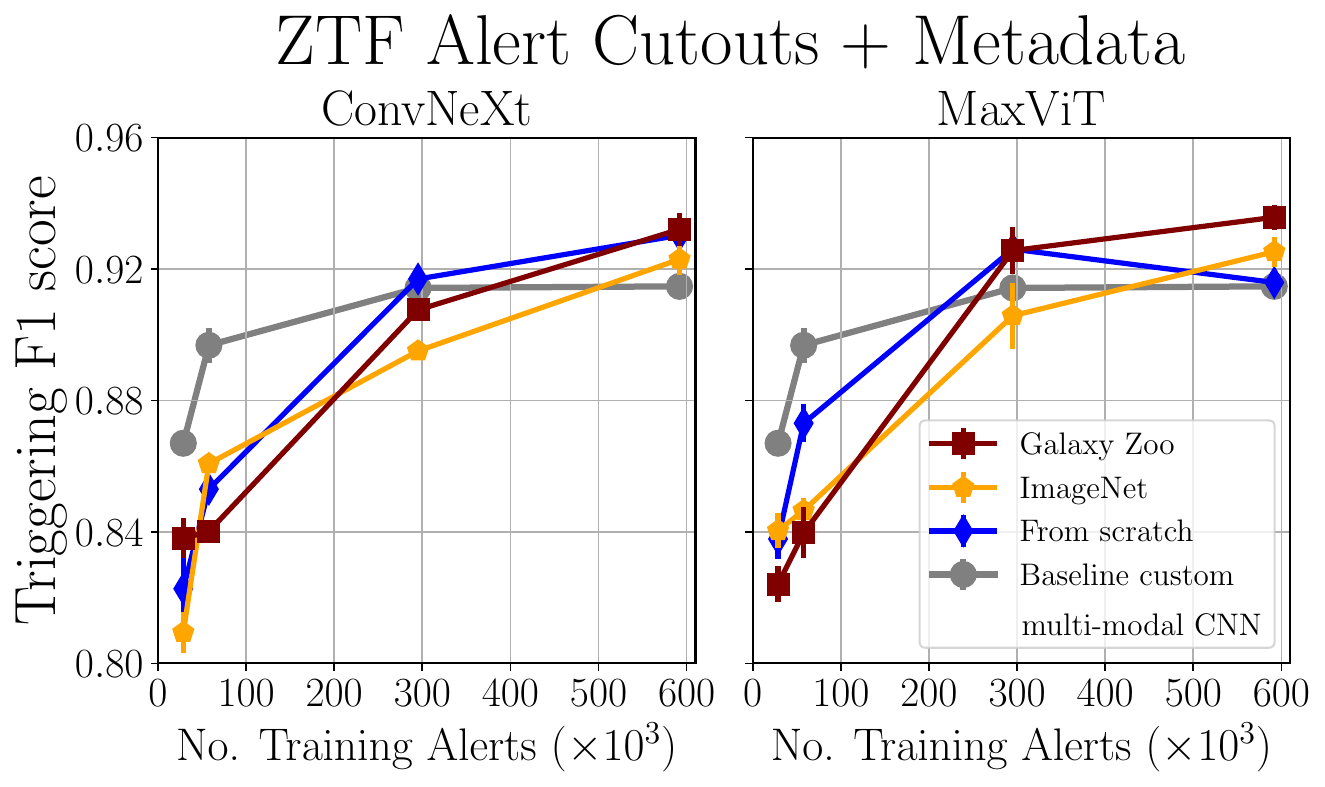}
    \caption{F1 scores representing how well multi-modal architectures with varying pre-training regimens can identify bright transient candidates for automatic spectroscopic follow-up compared against the baseline production \texttt{BTSbot} model (gray). At small training set sizes ($<100\times10^3$ alerts), the baseline production model out performs the off-the-shelf architectures across all pre-training options (Galaxy Zoo, maroon; ImageNet, orange; no pre-training, blue). As training set size increases ($>300\times10^3$ alerts), off-the-shelf architectures begin to outperform the baseline custom CNN, likely a product of their greater learning capacity. Pre-training on Galaxy Zoo unlocks the greatest overall performance across both the ConvNeXt and MaxViT architectures.}
    \label{fig:multimodal}
\end{figure*}

We additionally present multi-modal models benchmarked on bright transient candidate vetting to assess if an alternative vision model can improve the in-production performance of \texttt{BTSbot}. The primary task of \texttt{BTSbot} in production is to identify and allocate spectroscopic follow-up resources to classify new bright transients and infant supernovae. \cite{Rehemtulla+2024_BTSbot, Rehemtulla+2025} measured model performance on this task with the completeness and purity of triggers, reflecting the importance of recovering all transients of interest without triggering follow-up for other sources. Maximizing general classification performance was the goal of Section~\ref{sec:results-sm} which led us to focus on ROC AUC, but here our focus is on in-production performance, so we adopt the F1 score of triggers sent. An F1 score is the harmonic mean of the purity and completeness, so it summarizes the performance of both priorities in a single metric. Table~6 of \cite{Rehemtulla+2024_BTSbot} demonstrated that generic metrics like ROC AUC are not always directly correlated with in-production metrics, so by adopting this custom metric, we can most realistically assess the performance of a model deployed in production for automatically allocating follow-up resources. \revision{\cite{Rehemtulla+2024_BTSbot} introduces the concept of a \textit{policy}, a set of criteria which are used to determine when a request for follow-up is sent. Here, we use the F1 score of the \texttt{bts\_p1} policy which requires a source to have at least two alerts with \texttt{BTSbot} score $\geq$0.5 and at least one alert with $\texttt{magpsf}\leq19$ mag.\footnote{\revision{We note that slightly ($\sim1\%$) higher F1 scores can be achieved by increasing the score threshold from 0.5 to $\sim$0.8. As such, a policy using a higher score threshold is deployed in one of the real \texttt{BTSbot} production workflows. This gain is small, however, so we adopt the original \texttt{bts\_p1} policy here to maintain a direct comparison with the results presented in \cite{Rehemtulla+2024_BTSbot} without detracting from the relevance to production workflows.}}}

We also do not present multi-modal results using the Legacy Survey images because they do not represent a realistic production scenario: the mismatched sky coverage of ZTF and Legacy Survey would prevent classifications for many alerts (see Section~\ref{sec:choice-btsbot}), resulting in very poor completeness. Image cutouts from Pan-STARRS could be substituted in place of Legacy Survey to better match the sky coverage of ZTF, however, we do also observe substantially worse maximum performance from Legacy Survey cutouts than from ZTF alert cutouts so we do not explore this option. We do note, however, that this result can vary based on the task chosen to benchmark on.

Figure~\ref{fig:multimodal} reports the F1 score for triggering follow-up as a function of the number of alerts trained on. We find that, for both off-the-shelf architectures, the baseline yields better performance at small training set sizes ($\lesssim100,000$ alerts). For larger training sets sizes, the off-the-shelf architectures mostly match or surpass the performance of the baseline multi-modal CNN. The baseline ceases to improve from $\sim$300,000 to $\sim$600,000 total training alerts. We hypothesize this is the case because the small image branch does not have a sufficient learning capacity to take advantage of the increased training set size. This interpretation is consistent with the steadily positive sloped performance of the larger, off-the-shelf architectures.\footnote{The MaxViT model trained from scratch on $\sim$600,000 alerts is an exception to this trend. This is likely a symptom of its hyperparameter sweep not being broad enough or run for long enough to identify sufficiently performant hyperparameters (see Section~\ref{sec:hpo}).} 

The performance trends across pre-training regimens for multi-modal models are generally not as clear as those of the models trained with images only (Section~\ref{sec:results-sm}), but we do find that the relative performance difference between pre-trained models and models trained from scratch is clearly decreased. We attribute this to the overall reduced importance of the vision branch to this multi-modal prediction. \cite{Rehemtulla+2024_BTSbot} show that a model performing this task with the metadata features alone yields comparable performance to the multi-modal model on similar in-production metrics. Still, we do find that the models with vision branches pre-trained on Galaxy Zoo generally outperform those pre-trained with ImageNet. This again demonstrates that pre-training on Galaxy Zoo is more conducive to later fine-tuning on ZTF alerts than pre-training on ImageNet.

\section{Further model comparisons}
\label{sec:further_comparison}

\begin{figure*}[ht!]
    \centering
    \includegraphics[width=1\linewidth]{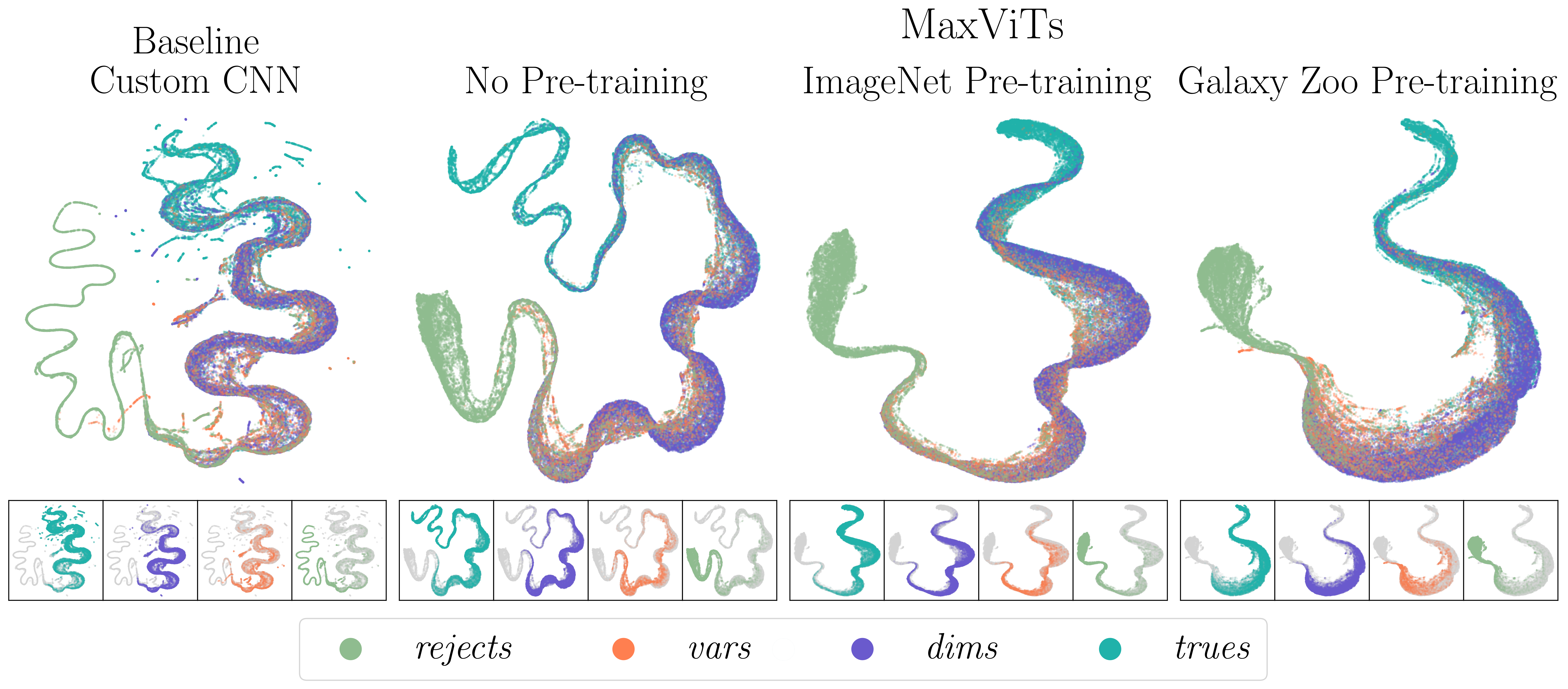}
    \caption{UMAP embeddings for the baseline custom CNN and MaxViT models with two pre-training regimens and another MaxViT trained from scratch. Points are colored by the subset of the training data from which they originate: \textit{trues} (teal) bright extragalactic transients; \textit{dims} (dark blue) faint sources with transient-like light curves; \textit{vars} (orange) AGN, CVs, quasars, VarStars; \textit{rejects} (light green) miscellaneous sources rejected as bright transient candidates. All latent spaces show the similar qualitative structure with a continuum having \textit{trues} and \textit{rejects} at opposite ends with \textit{dims} and \textit{vars} intermediate to them, but none are identical implying that the pre-training has a lasting effect on the models after fine-tuning.}
    \label{fig:umap}
\end{figure*}

Although performance metrics such as ROC AUC and our tailored in-production metrics remain essential for evaluating classification quality, they provide only a partial view of how models differ from one another. These metrics neglect the computational costs and can obscure more nuanced information on what the model has learned. In practice, two models with nearly identical ROC curves may impose vastly different computational burdens, exhibit distinct representational structures in their latent spaces, or behave differently under distribution shifts. For these reasons, we complement traditional performance metrics with analyses of computational efficiency (inference runtime and memory demands) and qualitative comparisons of the models’ learned feature spaces via uniform manifold approximation and projection \citep[UMAP; ][]{McInnes+2018} embeddings. These additional analyses offer a more holistic view of their trade-offs and practical suitability for real-time production environments of large image-differencing surveys.

\subsection{Computational cost}
\label{sec:computational}

\begin{deluxetable}{c|ccc}
\tablecaption{Computational cost of vision models \label{tab:comp}}
\tablehead{
    \colhead{\makecell{Architecture}} & \colhead{\makecell{Throughput \\ GPU [alert/s]}} & \colhead{\makecell{Throughput \\ CPU [alert/s]}} & \colhead{\makecell{Memory \\ $\textrm{[MB]}$}}
}
\startdata
    Custom CNN & 3708.5 & 136.1 & 345.5 \\
    ConvNeXt & \textbf{6622.3} & \textbf{874.6} & \textbf{62.7} \\
    MaxViT & 459.0 & 43.2 & 1045.6 \\
\enddata
\end{deluxetable}

To describe a model's computational cost, we evaluate the speed at which a model runs inference and the memory requirements to do so. These are critical considerations as the computer systems running survey pipelines have limited computational resources and, in certain scenarios, models can bottleneck the speed at which alerts are relayed to the community, adversely affecting time-sensitive science cases (e.g., multi-messenger astronomy). Furthermore, vision models are typically trained and designed to run on graphics processing units (GPUs) which very efficiently handle image inputs. However, most survey computer systems are not equipped with GPUs and are thus forced to run these vision models on central processing units (CPUs). We report runtime measurements as the image throughput for each model architecture running on both a GPU and a CPU. Throughput is the number of alerts in the test split divided by the time it takes to compute predictions for each of them using a batch size of 32 and 32-bit floating point precision. The GPU tests use an NVIDIA A30 GPU and the CPU tests use 12 cores of an Intel Xeon Gold 6338 CPU. The memory usage is measured as the peak GPU memory usage during the inference test. The numbers we report are the medians of three trials.

Table~\ref{tab:comp} shows the results of the computational cost analysis. \revision{The MaxViT shows a large computational burden: 20.2$\times$ and 3.2$\times$ slower on CPU throughput than the ConvNeXt and the custom CNN, respectively, while also requiring 16.7$\times$ and 3.0$\times$ more memory.} This is a product of the attention mechanisms used in the model which introduce large intermediate feature maps whose memory footprint scales roughly quadratically with the spatial dimensions of the input \citep{Dosovitskiy+2020, Tu+2022}. These costs are intrinsic to the MaxViT design as it is predominantly driven by these costly attention layers. The ConvNeXt achieves the best throughput and memory efficiency, despite having more trainable parameters than the custom CNN. Quantitatively, \revision{the ConvNeXt is 6.4$\times$ faster on CPU throughput and uses 5.5$\times$ less memory.} The cost advantage of the ConvNeXt stems from numerous factors of its design; here we highlight two. The first is that the ConvNeXt adopts ``depthwise convolutions" in which each channel input into a layer is convolved independently with its own kernel, rather than all channels being jointly processed as in a traditional convolutional layer. This is followed by a ``pointwise convolution" which uses a $1\times1$ kernel to restore the ability to mix information across channels which was prohibited by the depthwise convolution \citep{Chollet+2016, Howard+2017}. This dramatically reduces the total number of operations computed relative to an analogous standard convolutional layer. The second advantage is that ConvNeXt also adopts a large kernel stride in early convolutional layers which significantly decreases the spatial resolution of the input early on in the model. This has the effect of reducing the size of the intermediate feature maps produced within the model, reducing computation and memory requirements in later stages of the model. The custom CNN lacks these optimizations. Although it could be redesigned to adopt some of them, this illustrates one critical advantage of using off-the-shelf architectures; the design of a model architecture involves highly technical and nuanced choices which can be easily overlooked while still having a strong influence on the model's computational cost.

\subsection{Qualitative comparison of UMAP embeddings}
\label{sec:umap}

We explore 2-dimensional UMAP embeddings from MaxViT models to investigate whether or not models with distinct pre-training regimens converge to identical final states during fine-tuning. We extract high-dimensional representations from the penultimate layer of the models and fit a UMAP transformation to reduce their dimensionality and assist with making a human-interpretable visualization of the model's latent space. \revision{In all cases, we adopt default values for the UMAP transformation hyperparameters established in \cite{McInnes+2018} and the UMAP Python package:\footnote{\url{https://github.com/lmcinnes/umap}} $N_\textrm{neighbors}=15$, $\textrm{min\_dist}=0.1$, and $N_\textrm{components}=2$.} In Figure~\ref{fig:umap}, we show the UMAP embeddings for all alerts in the test split from the baseline custom CNN and three MaxViT models with different pre-training regimens or without pre-training. The points are colored by the subset of the training data from which they originate: \textit{trues} represents spectroscopically confirmed bright ($m_\mathrm{peak}\,\leq\,18.5\,\mathrm{mag}$) extragalactic transients; \textit{dims} represents dim ($m_\mathrm{peak}\,>\,18.5\,\mathrm{mag}$) sources with transient-like light curves; \textit{vars} represents sources classified as AGN, cataclysmic variables (CVs), VarStars, or quasars; and \textit{rejects} represents ZTF sources which pass the BTS alert filter but were not cataloged as bright extragalactic transients (see \citealt{Rehemtulla+2024_BTSbot} for details). We find that all four models have qualitatively similar latent spaces. They all place points along a continuum with one end nearly entirely populated by the bright extragalactic transients and the other nearly entirely populated by the \textit{rejects}. Intermediate to these, we observe significant overlap between the tail of the distribution of the \textit{trues} and the \textit{dims}. Many of the \textit{trues} alerts in this space do belong to bright transients but occur while the transient is fainter than the BTS threshold (i.e., $m\,>\,18.5\,\mathrm{mag}$). Overlapping the other end of the \textit{dims} are the \textit{vars} and some of the \textit{rejects}. This aligns with the overlap in source populations in these subsets because all three are expected to include, e.g., CVs and AGN flares. Beyond strong high-level similarities, the UMAP visualizations of the MaxViT latent spaces do show some local substructure differences. The Galaxy Zoo pre-trained MaxViT, e.g., exhibits peninsulas protruding from the main continuum while the others do not. We caution the over-interpretation of the UMAP visualizations as they represent only a 2-dimensional view of a much higher dimensional space, although we do find that these are suggestive that the pre-training has a lasting effect after the fine-tuning process.

We also explore the utility of UMAP embeddings for assisting with identifying outliers in the training set. We run an Isolation Forest \citep[IF;][]{Liu+2008} independently on each subset (i.e., one IF each for the \textit{trues}, \textit{vars}, \textit{dims} and \textit{rejects}) of test split for each MaxViT UMAP embedding and select the 1\% of alerts with the highest IF scores. We then select only the alerts which were marked as anomalous by all three MaxViTs. We visually inspect these sources and identify a number of interesting outliers. Some sources are genuinely rare and astrophysically interesting, e.g., ZTF21abzithf which is an instance of two Type Ia supernovae (SN\,2021yhz and SN\,2024svu) occurring at the same location.\footnote{One half of the alerts marked as anomalous from this source come from SN\,2024svu and the other half come from SN\,2021yhz.} Type Ia supernova ``siblings'' like these can be excellent aids for improving standardization relations in their use as cosmological probes \citep{Graham+2022, Dhawan+2025}. Other outliers are rare but less astrophysically interesting, e.g., ZTF20aaewhin (AT\,2017foz) which is a CV projected atop a background galaxy creating an image which looks much like a supernova. We also identify multiple supernovae with peak magnitudes very close to the BTS threshold ($18.5 < m_\textrm{peak} \textrm{ [mag]} < 18.6$), e.g. ZTF20aaetirf (SN\,2020mbe) with $m_\textrm{peak}=18.517$ mag. Lastly, we identify two sources with incorrect labels in our training set. ZTF19aalbaff (AT\,2019blg) is very likely a real extragalactic transient that was internally cataloged as an AGN because it is near the nucleus of its AGN host. ZTF22aafvbku is a duplicate ZTF source of ZTF22aadfrjm (SN\,2022fbf) which is a Type Ia supernova (i.e., both contain mostly the same supernova light curve). The incorrect label originated from this duplicate ZTF source being rejected from BTS instead of being cataloged as a bright transient alongside its copy. Duplicates are rarely rejected instead of being cataloged, so we do not expect there is a significant population of other sources mislabeled in this way.

These findings demonstrate that pairing UMAP embeddings with IF enables the recovery of astrophysically interesting sources as well as label noise in our training set. Further use of UMAP and IF on \texttt{BTSbot} embeddings could assist with the identification of bright transients missed by the BTS team, a critical factor in their supernova rates calculations.

\section{Discussion and Conclusions} \label{sec:discussion}

Our experiments reveal several distinct advantages of using pre-trained, off-the-shelf models for alert classification. Compared to a baseline model specially designed for this transient candidate vetting task, these models consistently achieve equivalent or superior performance across a wide range of training set sizes. From another lens, the pre-trained off-the-shelf models are able to attain an equal level of performance with as many or fewer labeled examples. Off-the-shelf architectures also yield higher maximum performance than the baseline model (Section~\ref{sec:results-sm}, \ref{sec:results-mm}). We also find that pre-training always improves performance relative to training from scratch for the models using only images for prediction. Critically, pre-training on Galaxy Zoo generally facilitates better performance than pre-training on ImageNet. We argue that this finding is non-trivial as there remain key differences between the Galaxy Zoo and ZTF alert cutouts datasets. (i) ZTF alert cutouts are not red-green-blue color images like the Galaxy Zoo pre-training data; (ii) Galaxy Zoo images always have a galaxy centered in the cutout, while ZTF cutouts often have off-center galaxies or no galaxies at all; and (iii) the Galaxy Zoo data come from observatories with different noise and detector properties than ZTF images. 

The increases in performance metrics are modest in absolute terms, but they are made to models that already deliver a high level of performance. Furthermore, these modest gains have meaningful scientific impact when (i) data volumes scale significantly, as expected by LSST, and (ii) when models are used for tasks with minimal tolerance for misclassifications, such as autonomously triggering target of opportunity requests to oversubscribed facilities \citep{Rehemtulla+2025}. 

Beyond performance metrics, off-the-shelf model architectures offer substantial benefits in terms of computational efficiency and ease-of-use. The process of designing, debugging and perfecting a custom model architecture can take many months, but integrating a publicly available, meticulously-designed, and standard architecture into one's code base takes just a small fraction of that time. Off-the-shelf architectures, like ConvNeXt are designed to be highly efficient (see Section~\ref{sec:further_comparison}). Nuanced choices made in the ConvNeXt design yield throughput and memory efficiency considerably better than the smaller, custom CNN which is representative of models presented in the literature. These efficiency gains are especially pertinent in the present day, as new image-differencing surveys are quickly coming online and alert rates are growing. New models must be rapidly transferred across surveys with minimal labeled data to aid with quickly beginning productive science operations.

Given these demonstrated benefits, we advocate for the broader adoption of pre-trained off-the-shelf models for classifying alerts from image-differencing surveys. The strong performance and low computational cost of the ConvNeXt models indicate they are an excellent choice for the bulk of applications. The MaxViT models do unlock higher maximum performance, so practical deployments may benefit from hybrid configurations, where attention-based models are used for high-priority or ambiguous alerts, while convolutional models process the bulk of the stream. 

To facilitate this transition, we release \texttt{BTSbot} models presented here publicly on the \texttt{Hugging Face} platform,\footnote{\url{https://huggingface.co/collections/nabeelr/pre-trained-off-the-shelf-btsbots}} along with open-source training code on GitHub.\footnote{\url{https://github.com/nabeelre/BTSbot}} We intend for these models to be used as a strong starting point for fine-tuning to other alert classification tasks on ZTF alerts or those from another survey. Lastly, we are deploying the multi-modal ConvNeXt pre-trained on Galaxy Zoo into production to assume the operational roles of the current \texttt{BTSbot} model. The considerably higher throughput of the ConvNeXt vision backbone will prevent \texttt{BTSbot} inference from being a bottleneck in the alert processing pipelines of the brokers it is deployed in. The throughput of the custom CNN demonstrated with 12 CPU cores (see Section~\ref{sec:computational}) is insufficient to keep up with the LSST alert stream which is expected to average upwards of $\mathcal{O}(10^2)$ alerts per second. Furthermore, the throughput of the larger, multi-modal production \texttt{BTSbot} is yet smaller. Thus, these updates to the \texttt{BTSbot} model architecture support the model's longevity in the increasingly demanding landscape of real-time production environments and mark a transition toward processing alerts with pre-trained off-the-shelf vision models.

\section{Acknowledgments}

We thank members of the Northwestern Image and Video Processing Lab (Patrick Koller, Santiago L\'opez-Tapia, Philipp Srivastava) and the Bio Inspired Vision Lab (Emma Alexander, Mehmet Kerem Aydin, Tianao Li) for insightful discussions. We also thank the ZTF ML technical working group for helpful comments.

N.R. is supported by a Northwestern University Presidential Fellowship Award. N.R. and A.A.M.~are partially supported by DoE award \#DE-SC0025599. A.A.M.~is also supported by Cottrell Scholar Award \#CS-CSA-2025-059 from the Research Corporation for Science Advancement. N.R.~is also partially supported by NSF grant \# AST-2421845.

We gratefully acknowledge the support of the NSF-Simons AI-Institute for the Sky (SkAI) via grants NSF AST-2421845 and Simons Foundation MPS-AI-00010513.

Zwicky Transient Facility access for N.R., A.A.M. and V.G.S. was supported by Northwestern University and the Center for Interdisciplinary Exploration and Research in Astrophysics (CIERA).

M.~W.~Coughlin acknowledges support from the National Science Foundation with grant numbers PHY-2409481, PHY-2308862 and PHY-2117997.

This research was supported in part through the computational resources and staff contributions provided for the Quest high performance computing facility at Northwestern University which is jointly supported by the Office of the Provost, the Office for Research, and Northwestern University Information Technology.

The Gordon and Betty Moore Foundation, through both the Data-Driven Investigator Program and a dedicated grant, provided critical funding for SkyPortal. This publication uses data generated via the Zooniverse.org platform, development of which is funded by generous support, including a Global Impact Award from Google, and by a grant from the Alfred P. Sloan Foundation.

Based on observations obtained with the Samuel Oschin Telescope 48-inch and the 60-inch Telescope at the Palomar Observatory as part of the Zwicky Transient Facility project. ZTF is supported by the National Science Foundation under Grants No. AST-1440341, AST-2034437, and currently Award \#2407588. ZTF receives additional funding from the ZTF partnership. Current members include Caltech, USA; Caltech/IPAC, USA; University of Maryland, USA; University of California, Berkeley, USA; University of Wisconsin at Milwaukee, USA; Cornell University, USA; Drexel University, USA; University of North Carolina at Chapel Hill, USA; Institute of Science and Technology, Austria; National Central University, Taiwan, and OKC, University of Stockholm, Sweden. Operations are conducted by Caltech's Optical Observatory (COO), Caltech/IPAC, and the University of Washington at Seattle, USA.

This research has made use of NASA's Astrophysics Data System.


%

\vspace{5mm}
\facilities{PO:1.2m}


\software{\texttt{astropy} \citep{astropy:2013, astropy:2018, astropy:2022}, \texttt{matplotlib} \citep{hunter07}, \texttt{numpy} \citep{numpy}, \texttt{pandas} \citep{pandas1, pandas2}, \texttt{penquins} \citep{penquins}, \texttt{scikit-learn} \citep{sklearn}, \texttt{PyTorch} \citep{Paszke+2019}, \texttt{SkyPortal} \citep{van_der_Walt+2019, Coughlin+2023}, \texttt{timm} \citep{timm}, and the Weights and Biases platform \citep{wandb}}





\bibliography{tda_refs}{}
\bibliographystyle{aasjournal}



\end{document}